\documentclass[showpacs,10pt,twocolumn,prb]{revtex4-1}

\usepackage{amsmath}
\usepackage{amssymb}
\usepackage{graphicx}
\usepackage{CJK}
\usepackage{color}

\begin{document}

\title{Physical properties of K$_{x}$Ni$_{2-y}$Se$_{2}$ single crystals}
\author{Hechang Lei,$^{1}$ Kefeng Wang,$^{1}$ Hyejin Ryu,$^{1,2}$ D. Graf,$^{3}$ J. B. Warren,$^{4}$ and C. Petrovic$^{1,2}$}
\affiliation{$^{1}$Condensed Matter Physics and Materials Science Department, Brookhaven National Laboratory, Uptown, NY 11973 USA\\
$^{2}$Department of Physics and Astronomy, State University of New York, Stony Brook, NY 11794-3800, USA\\
$^{3}$NHMFL/Physics, Florida State University, Tallahassee, Florida 32310, USA\\
$^{4}$Instrumentation Division, Brookhaven National Laboratory, Upton, New
York 11973, USA}
\date{\today}

\begin{abstract}
We have synthesized K$_{0.95(1)}$Ni$_{1.86(2)}$Se$_{2}$ single crystals. The single crystals contain K and Ni deficiencies not observed in KNi$_{2}$Se$_{2}$ polycrystals. Unlike KNi$_{2}$Se$_{2}$ polycrystals, the superconductivity is absent in single crystals. The detailed physical property study indicates that the K$_{0.95}$Ni$_{1.86}$Se$_{2}$ single crystals exhibit Fermi liquid behavior with heavy-fermion-like characteristics. Transition from the high temperature local charge density wave state to heavy fermion state below $T\sim$ 30 K results in an enhancement of electron-like carrier density whereas magnetic susceptibility shows little anisotropy and suggests presence of both itinerant and localized Ni orbitals.
\end{abstract}

\maketitle

\section{Introduction}
Even before the discovery of superconducting $T_{c}$ = 26 K in LaFeAsO$_{1-x}$F$_{x}$,\cite{Kamihara} some nickel pnictide materials (such as LaONiP)\cite{Watanabe} have already been found to become superconducting (SC) at low temperature. Examples of nickel pnictide superconductors include also LaONiAs, LaO$_{1-x}$NiBi, BaNi$_{2}$P$_{2}$ and SrNi$_{2}$P$_{2}$.\cite{Watanabe2,Kozhevnikov,Mine,Ronning1} However, all nickel pnictide SCs have much lower $T_{c}$ ($<$ 5 K) when compared to iron pnictide SCs of which $T_{c}$'s are mostly well above 5 K.\cite{Ronning} The possible reason leading to this difference could be due to the different superconducting mechanisms or to the different values of materials parameters relevant for superconductivity which are not optimized for the nickel pnictides even if the pairing mechanism would be identical.\cite{Ronning}

On the other hand, iron chalcogenide SCs have also been discovered, including FeCh (Ch = S, Se, Te),\cite{Hsu FC}$^{-}$\cite{Mizuguchi} and A$_{x}$Fe$_{2-y}$Se$_{2}$ (A = K, Rb, Cs, Tl/K and Tl/Rb).\cite{Guo}$^{-}$\cite{Fang MH} In contrast to iron chalcogenide SCs, however, corresponding nickel chalcogenide SCs are still missing. NiSe has NiAs type structure with space group P63/mmc and this structure is isostrucutral to hexagonal FeSe (high temperature phase). NiSe is a non-superconducting metal with ferromagnetic fluctuations.\cite{Umeyama} Similarly, TlNi$_{2}$Se$_{2}$ is a Pauli paramagnet without superconducting transition down to 2 K.\cite{Newmark} However, very recently, it was reported that KNi$_{2}$Se$_{2}$ polycrystals are superconducting with $T_{c}$ = 0.80(1) K.\cite{Neilson2} Moreover, KNi$_{2}$Se$_{2}$ single crystals feature a heavy-fermion-like state with an increased carrier mobility and enhanced effective electronic band mass below about 40 K. This state should emerge from the local charge density wave (CDW) state which persists up to 300 K.

A study of single crystals is necessary in order to elucidate the anisotropy in intrinsic physical properties of KNi$_{2}$Se$_{2}$ and eliminate the influence of grain boundaries and ferromagnetic Ni impurities. Hence in this work we report the physical properties of K$_{0.95(1))}$Ni$_{1.86(2)}$Se$_{2.00(1)}$ single crystals. Unexpectedly, we found no evidence for superconducting transition down to 0.3 K in resistivity measurement. Magnetic, Hall, thermodynamic and thermal measurements suggest that the heavy-fermion-like properties of K$_{x}$Ni$_{2-y}$Se$_{2}$ arise from dominant electron-like carriers. The Fermi surface reconstruction with increased electron-like carrier concentration at $T\sim$ 30 K marks the crossover from local CDW state at high temperatures to the low temperature heavy-fermion-like state.

\section{Experiment}
Single crystals of KNi$_{2}$Se$_{2}$ were grown by self-flux method similar to K$_{x}$Fe$_{2-y}$Se$_{2}$\cite{Lei HC2} with nominal composition K:Ni:Se = 1:2:2. Briefly, prereacted NiSe and K pieces
were added into the alumina crucible with partial pressure of argon gas. The quartz tubes were heated to 1030 ${^{\circ }}C$, kept at this temperature for 3 hours, then cooled to 730 ${^{\circ }}C$
with 6 ${^{\circ }}C$/h. The platelike dark pink colored single crystals with typical size 5$\times$5$\times $2 mm$^{3}$ can be grown. X-ray diffraction (XRD) spectra were taken with Cu K$_{\alpha}$
radiation ($\lambda $=$1.5418$ \r{A}) using a Rigaku Miniflex X-ray machine. The lattice parameters were obtained by fitting the XRD spectra using the Rietica software.\cite{Hunter} The elemental
analysis was performed using an energy-dispersive x-ray spectroscopy (EDX) in an JEOL JSM-6500 scanning electron microscope. Electrical and thermal transport, heat capacity, and magnetization
measurements were carried out in Quantum Design PPMS-9 and MPMS-XL5. The in-plane resistivity $\rho _{ab}(T)$ was measured using a four-probe
configuration on cleaved rectangularly shaped single crystals with current flowing in the $ab$-plane of tetragonal structure. 
Thin Pt wires were attached to electrical contacts made of silver paste. Thermal transport properties were measured in Quantum Design PPMS-9 from 2 to 350 K using a one-heater two-thermometer method. The relative error was $\frac{\Delta \kappa}{\kappa}\sim$5$\%$ and $\frac{\Delta S}{S}\sim$5$\%$ based on Ni standard measured under identical conditions.

\section{Results and Discussion}
As shown in Fig. 1(a), powder XRD pattern of KNi$_{2}$Se$_{2}$ can be fitted very well using the I4/mmm space group. The determined lattice parameters are $a$ = 0.3899(2) nm and $c$ = 1.3473(2) nm. The value for $a$ axis is somewhat smaller wheras the $c$ axis lattice parameter is larger when compared to polycrystalline samples with full occupancies of K and Ni ($a$ = 0.39098(8) nm and $c$ = 1.34142(5) nm.\cite{Neilson1} This is in agreement with previous study which indicated that the deviation from full occupancy can increase the $c$ lattice parameter with only minor effect on $a$ lattice parameter.\cite{Neilson1} On the other hand, both $a$ and $c$ lattice parameters are smaller than those in K$_{x}$Fe$_{2-y}$Se$_{2}$ (a = 0.39109 nm and c = 1.4075 nm).\cite{Lei HC2} It could be due to the smaller ionic radius of Ni$^{2+}$ (55 pm) than Fe$^{2+}$ (63 pm) with four-fold coordination.\cite{Speight} The antiferromangetic state in  K$_{x}$Fe$_{2-y}$Se$_{2}$ can increase the lattice parameters further when compared to the non-magnetic state according to the theoretical calculation.\cite{Cao C} The crystal structure of KNi$_{2}$Se$_{2}$ is shown in Fig. 1(b), where antifluorite-type Ni-Se layers and K cation layers are piled up alternatively along the $c$ axis. XRD pattern of a single crystal (Fig. 1(c)) reveals that the crystal surface is normal to the $c$ axis with the plate-shaped surface parallel to the $ab$-plane. Fig. 1(d) shows the EDX spectrum of a single crystal of KNi$_{2}$Se$_{2}$, confirming the presence of the K, Ni, and Se. The EDX results for several single crystals with multiple measuring points indicate that the crystals are rather homogenous and the determined average atomic ratios are K:Ni:Se = 0.95(1):1.86(2):2.00(1) when fixing Se stoichiometry to be 2. Although the amount of deficiencies is smaller when compared to K$_{x}$Fe$_{2-y}$Se$_{2}$ crystals,\cite{Lei HC2} it is noticeably higher when compared to polycrystalls which feature full occupancies of K, Fe and Se atomic sites.

\begin{figure}[tbp]
\centerline{\includegraphics[scale=0.18]{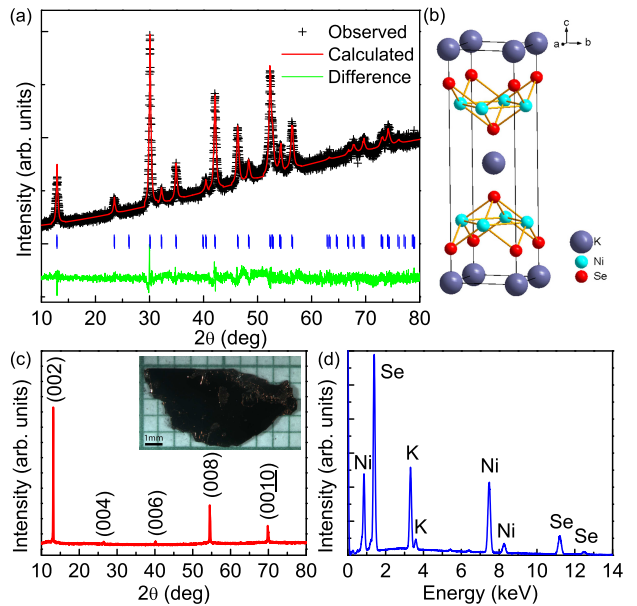}} \vspace*{-0.3cm}
\caption{(a) Powder XRD patterns of K$_{0.95}$Ni$_{1.86}$Se$_{2}$. (b) Crystal structure of KNi$_{2}$Se$_{2}$. The big purple, medium cyan and small red balls represent K, Ni, and Se ions. (c) Single crystal XRD of K$_{0.948}$Ni$_{1.86}$Se$_{2}$. The inset shows a photo of typical single crystal of KNi$_{2}$Se$_{2}$. The crystals are less air sensitive when compared to K$_{x}$Fe$_{2-y}$Se$_{2}$ single crystals. (d) The EDX spectrum of a single crystal.}
\end{figure}

\begin{figure}[tbp]
\centerline{\includegraphics[scale=0.35]{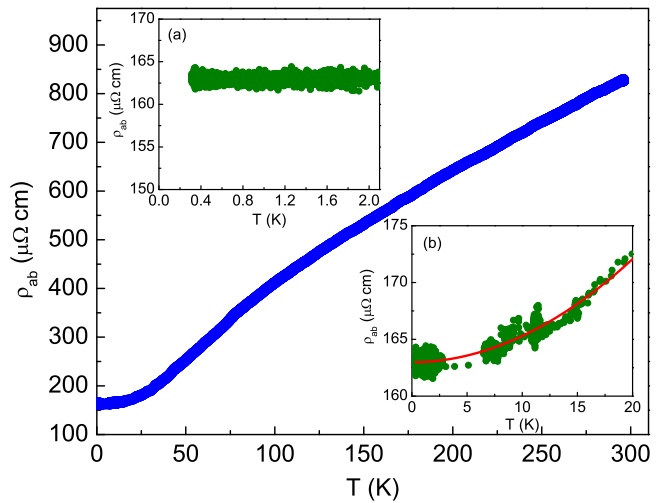}} \vspace*{-0.3cm}
\caption{Temperature dependence of the in-plane resistivity $\rho_{ab}(T)$ of the K$_{0.95}$Ni$_{1.86}$Se$_{2}$ single crystal from 0.3 K to 295 K. The inset (a) shows the enlarged part of $\rho_{ab}(T)$ below 2 K. The inset (b) shows the fitted result from 0.3 K to 20 K using $\rho_{ab}(T) =\rho _{0}+AT^n$ where the red line is the fitting curve.}
\end{figure}

The temperature dependence of the in-plane resistivity $\rho_{ab}(T)$ of the K$_{0.95}$Ni$_{1.86}$Se$_{2}$ single crystal exhibits metallic behavior with the residual resistivity ratio (RRR) $\rho_{ab}$(295 K)/$\rho_{ab}$(0.3 K) = 5. The room-temperature value $\rho _{ab}$(295 K) (256 $\mu\Omega \cdot $cm) is slightly smaller than the value in polycrystals (350 $\mu\Omega \cdot $cm). However, the RRR value is much smaller than that in polycrystals (RRR$\sim$175) possibly due to the absence or minimization of Ni impurity in single crystals (high purity Ni can have RRR$>$2000).\cite{Neilson2,Ehrlich} In addition, larger residual resistivity $\rho_{0}$ in crystals could come from the increased impurity scattering due to the deficiencies. The most striking feature is the absence of superconducting transition down to 0.3 K in K$_{0.948}$Ni$_{1.86}$Se$_{2}$ single crystals when compared to KNi$_{2}$Se$_{2}$ polycrystals with $T_{c}$ = 0.80(1) K.\cite{Neilson2} The absence of superconductivity in single crystal could be related to the deficiencies of K and Ni, implying that the superconductivity in KNi$_{2}$Se$_{2}$ is very sensitive to the atomic ratio. There is no metal-insulating transition (MIT) in KNi$_{2}$Se$_{2}$ and the absolute values of resistivity are much smaller when compared to K$_{x}$Fe$_{2-y}$Se$_{2}$,\cite{Lei HC2} indicating that Ni orbitals in the former are more itinerant when compared to Fe orbitals in the latter material. Surprisingly, as shown in the inset of Fig. 2(b), the $\rho_{ab}(T) \sim T^{2}$ dependence is observed up to 20 K at temperatures where other types of scattering (e.g. electron-phonon) are usually active or dominant. From the fit using $\rho_{ab}(T) =\rho _{0}+AT^n$, we obtain the residual resistivity $\rho_{0}$ = 50.282(2) $\mu\Omega $cm, the coefficient of the quadratic resistivity term $A$ = 0.0079(5)$\mu\Omega $cm/K$^{2}$, and $n$ = 1.96(2). It indicates that KNi$_{2}$Se$_{2}$ enters Fermi-liquid regime at low temperature.


\begin{figure}[tbp]
\centerline{\includegraphics[scale=0.38]{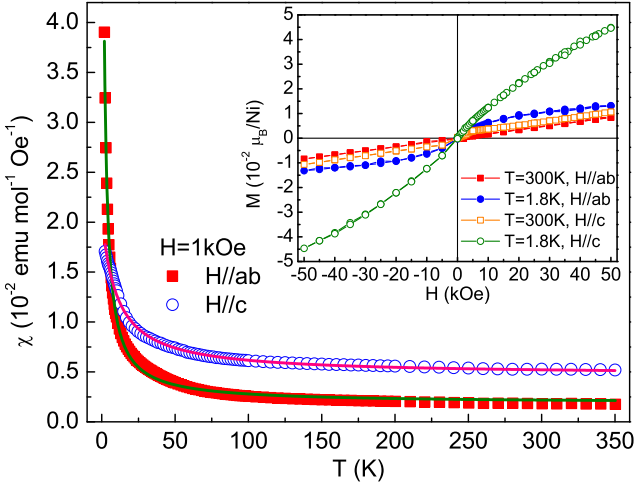}} \vspace*{-0.3cm}
\caption{(a) Temperature dependence of magnetic susceptibility $\chi $(T) with the applied field $H$ = 1 kOe along ab plane and c axis below 300 K. The inset shows the isothermal magnetization hysteresis loops M(H) for H$\Vert $ab and H$\Vert $c at 300 K and 1.8 K.}
\end{figure}

Fig. 3 presents the temperature dependence of magnetic susceptibility $\chi(T)$ of the K$_{0.95}$Ni$_{1.86}$Se$_{2}$ single crystal for $H$ = 1 kOe along the $ab$ plane and $c$ axis below 300 K. It can be seen that the $\chi _{c}(T)$ is larger than the $\chi _{ab}(T)$ at high temperature. The susceptibility can be fitted using Curie-Weiss law $\chi (T)=\chi _{0}+C/(T-\theta)$ when a temperature-independent contribution $\chi_{0}$ is accounted for. Here, $\chi_{0}$ includes core diamagnetism, Landau diamagnetism, and Pauli paramagnetism, $C$ is Curie constant and $\theta $ is the Curie-Weiss temperature (solid lines in Fig. 3). The fitted values for $\chi_{0}$ are 1.87(5)$\times$10$^{-3}$ and 4.69(4)$\times$10$^{-3}$ emu mol f.u.$^{-1}$ Oe$^{-1}$ for H$\Vert$ab and H$\Vert$c, which are much larger than the value in literature.\cite{Neilson1} Because the core diamagnetism is typically on the order of 10$^{-6}$-10$^{-5}$ emu mol$^{-1}$ Oe$^{-1}$,\cite{Carlin} and $\chi_{Landau}\approx-1/3\chi_{Pauli}$, such large $\chi_{0}$ values strongly imply there is an enhanced Pauli paramagnetism, i.e. the significant pileup of the density of states at the Fermi level due to $\chi_{Pauli}\approx\mu_{B}^{2}N(E_{F})$. On the other hand, the obtained local moment is about 0.463(3) and 0.615(7) $\mu_{B}$/Ni for H$\Vert$ab and H$\Vert$c. This is unlikely to be due to impurities such as Ni$^{2+}$ with S = 1 because the corresponding molar fraction would be 16.4(1) and 29.7(3) mol\% for H$\Vert$ab and H$\Vert$c, respectively. Such large amount of impurities should have been detected in the XRD pattern, hence the origin of low temperature susceptibility rise should be intrinsic. Contribution of impurity is revealed (Fig. 3 (inset)) in the magnetization loops for both field directions at 300 K and 1.8 K. There is a ferromagnetic component superposed on the paramagnetic background with very small magnetic moment ($\sim10^{-2} \mu_{B}$/Ni). After subtracting paramagnetic part from the curve for H$\Vert$ab at 1.8 K, the saturated moment would correspond to $\sim$0.2 mol\% Ni$^{2+}$ or $\sim$1 mol\% Ni impurities. The above analysis indicates that Ni orbitals in NiSe$_{4}$ tetrahedra are at the boundary of itinerancy and Mott localization with possible orbital dependent correlation strength, similarly to iron orbitals in iron based superconductors.\cite{YinZP}

\begin{figure}[tbp]
\centerline{\includegraphics[scale=0.35]{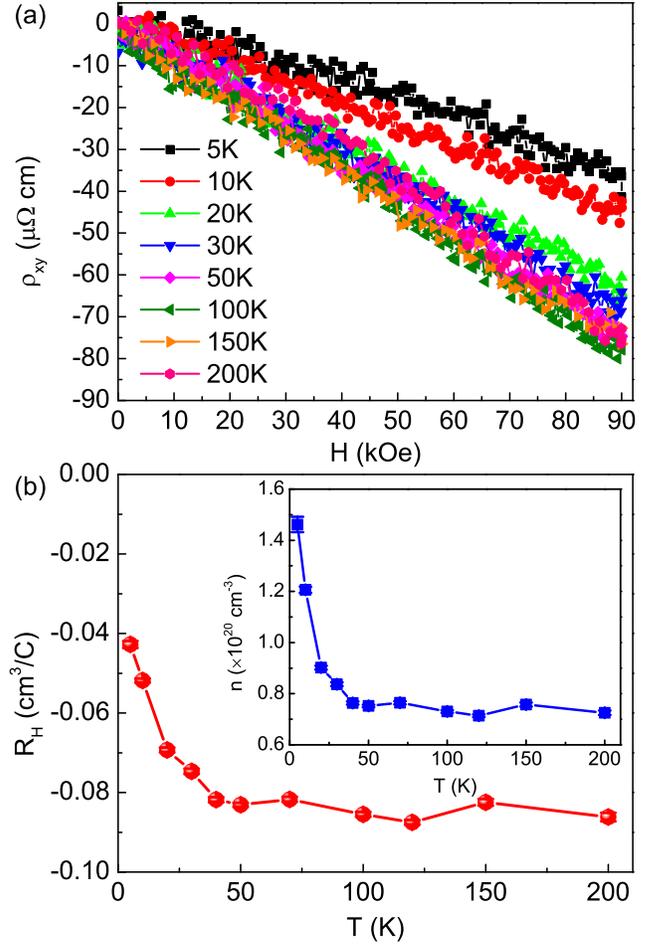}} \vspace*{-0.3cm}
\caption{(a) Field dependence of $\rho_{xy}(H)$ at various temperatures. (b) Temperature dependence of Hall coefficient $R_{H}$ of K$_{0.95}$Ni$_{1.86}$Se$_{2}$ single crystal. Inset: Temperature dependence of carrier density $n=1/|eR_{H}|$ calculated from $R_{H}$ based on the single band model.}
\end{figure}

As shown in Fig. 4(a), the transverse resistivity $\rho_{xy}(H)$ shows approximately linear relation against the magnetic field and is negative at all measuring temperatures, indicating that the electron-type carrier is dominant. From the linear fitting of $\rho_{xy}(H)-H$ relation, we obtain the Hall coefficients $R_{H}=\rho_{xy}(H)/H$ at different temperatures, which is shown in Fig. 4(b). It can be seen that the $R_{H}$ is weakly temperature dependent when $T>$ 30 K and then decreases with temperature. In a single band scenario, this change suggest that the carrier density increases at about $T$ = 30 K since $R_{H}=1/ne$ (Inset of Fig. 4(b)). This could be related to the crossover from local CDW state to heavy fermion state in KNi$_{2}$Se$_{2}$.\cite{Neilson2} On the other hand, this can also be ascribed to the multiband effect, which has been observed in classic two band materials such as MgB$_{2}$ as well as in iron based material Nd(O,F)FeAs.\cite{Yang H, Cheng P}  However, the observed temperature dependence is much weaker than that in Nd(O,F)FeAs single crystal, which exhibits significant multiband behavior. It implies the multiband behavior should be weaker in KNi$_{2}$Se$_{2}$ when compared to iron based superconductors. From obtained $R_{H}$, the corresponding carrier density at 300 K is $\sim$ 4$\times$10$^{-3}$ carrier per Ni, increasing up to $\sim$ 8$\times$10$^{-3}$ carrier per Ni. The carrier density is very low, even one order lower than in iron based superconductors, such as LaFeAsO$_{0.89}$F$_{0.11}$ ($\sim$ 10$^{21}$ cm$^{-3}$).\cite{Sefat} Moreover, from the measured resistivity at 5 K ($\rho_{ab}$(5 K) = 4.74$\times$10$^{-5}$ $\Omega$ cm) and derived carrier density at same temperature ($n$(5 K) = 1.46$\times$10$^{20}$ cm$^{-3}$), the carrier mobility at 5 K can be roughly estimated using $\sigma=ne\mu$ and is about 905 cm$^{2}$ V$^{-1}$ S$^{-1}$. It is close to the result derived from magnetoresistance measurement of polycrystals (1070 cm$^{2}$ V$^{-1}$ S$^{-1}$), implying that KNi$_{2}$Se$_{2}$ has high carrier mobility. On the other hand, as discussed below, the electron specific heat is large at low temperature, indicating the increased effective mass. Because $\mu$ is proportional to mean scattering time and inversely proportional to the effective mass, the mean scattering time could be rather large at low temperatures.\cite{Neilson2}

\begin{figure}[tbp]
\centerline{\includegraphics[scale=0.35]{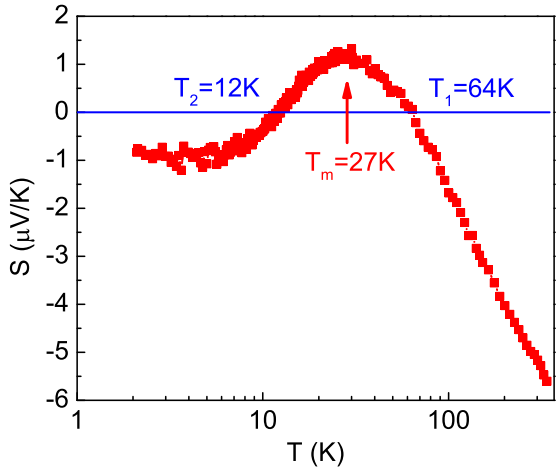}} \vspace*{-0.3cm}
\caption{Temperature dependence of thermoelectric power $S(T)$ for K$_{0.95}$Ni$_{1.86}$Se$_{2}$ single crystal from 2 to 350 K.}
\end{figure}

Fig. 5 shows the temperature dependence of thermoelectric power (TEP) $S(T)$ for K$_{0.95}$Ni$_{1.86}$Se$_{2}$ single crystals measured between $T$ = 2 and 350 K. At high temperature, the TEP is negative, consistent with dominant negative charge carriers. It is interesting that the value of the TEP decreases with decrease of temperature and then becomes positive at about $T_{1}$ = 64 K. The sign change implies multiband transport. Even though the Hall coefficient $R_{H}$ is unchanged in that temperature range, the sign of TEP $S(T)$ might change since they have different dependence on carrier density $n_{e}$ ($n_{h}$), mobility $\mu_{e}$ ($\mu_{h}$), and $S_{e}$ ($S_{h}$) in the two band model ($R_{H}=\frac{1}{e}\frac{n_{h}\mu_{h}^{2}-n_{e}\mu^{2}_{e}}{(n_{h}\mu_{h}+n_{e}\mu_{e})^{2}}$, $S=\frac{S_{e}n_{e}\mu_{e}+S_{h}n_{h}\mu_{h}}{n_{e}\mu_{e}+n_{h}\mu_{h}}$).\cite{Smith} With further decreasing temperature, TEP shows a peak at around $T_{m}$ = 27 K and then decreases with temperature. Finally it becomes negative again and the temperature corresponding to the second sign reverse is about $T_{2}$ = 12 K. The temperature of $T_{m}$ is very close to the temperature where the change of slope in $R_{H}$ appears. This suggests that the second sign change of $S(T)$ could be related to the crossover from local CDW state to heavy fermion state where Fermi surface becomes large with increasing carrier density in electron-type bands.

\begin{figure}[tbp]
\centerline{\includegraphics[scale=0.35]{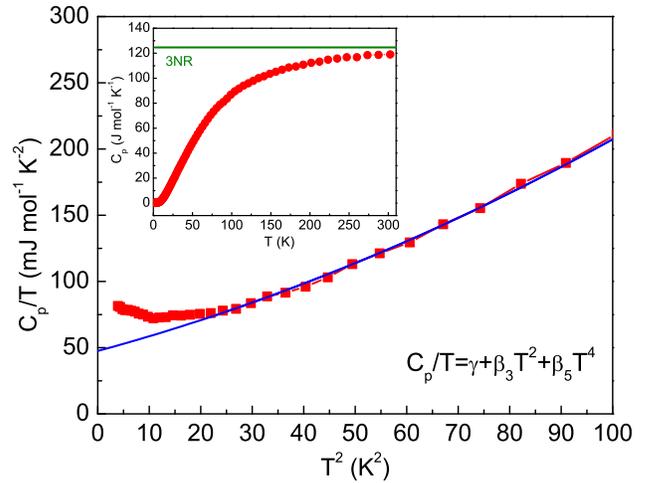}} \vspace*{-0.3cm}
\caption{The relation between $C_{p}/T$ and $T^{2}$ for K$_{0.948}$Ni$_{1.86}$Se$_{2}$ single crystal at low temperature. The solid curve represents the fitting result using the formula $C_{p}/T=\gamma+\beta_{3}T^{2}+\beta_{5}T^{4}$. The inset shows the temperature dependence of $C_{p}(T)$ for whole measuring temperature. The green solid line represents the classical value according to Dulong-Petit law at high temperature.}
\end{figure}

Fig. 6 shows the relation between $C_{p}/T$ and $T^{2}$ for K$_{0.95}$Ni$_{1.86}$Se$_{2}$ single crystal at low temperature. It can be seen that there is an upturn in the specific heat as $T<$ 3 K. This upturn could not be related to the superconducting transition or to nuclear Schottky anomaly since it appears at much higher temperatures. This might be an intrinsic effect related to magnetic fluctuations due to the deficiencies of K and Ni. Similar behavior has been observed in Ca(Fe$_{0.1}$Co$_{0.9}$)$_{2}$P$_{2}$ and Ca(Fe$_{1-x}$Ni$_{x}$)$_{2}$P$_{2}$ ($x$ = 0.5 and 0.6).\cite{Jia S}  In order to obtain the electronic specific heat and Debye temperature, we fit the $C_{p}/T-T^{2}$ curve between 5 and 10 K by using the formula $C_{p}/T=\gamma+\beta_{3}T^{2}+\beta_{5}T^{4}$. We obtain $\gamma$ = 48(2) mJ mol$^{-1}$ K$^{-2}$, $\beta_{3}$ = 1.06(8) mJ mol$^{-1}$ K$^{-4}$, and $\beta_{5}$ = 5.4(6) $\mu$J mol$^{-1}$ K$^{-6}$. The Debye temperature is estimated to be $\Theta _{D}$ = 209(5) K using the formula $\Theta _{D}$ = $(12\pi^{4}NR/5\beta )^{1/3}$, where N is the atomic number in the chemical formula (N = 5) and R is the gas constant (R = 8.314 J mol$^{-1} $ K$^{-1}$). The Debye temperature $\Theta _{D}$ and electronic specific heat $\gamma$ are close to the results obtained on the KNi$_{2}$Se$_{2}$ polycrystals.\cite{Neilson2} The large $\gamma$ implies the mass enhancement at low temperatures and heavy-fermion-like behavior.\cite{Neilson2} Moreover, the $\Theta _{D}$ is also similar to that of K$_{x}$Fe$_{2-y}$Se$_{2}$ ($\sim$ 212 K),\cite{Zeng B} which could be ascribed to the similar structure and atomic weight. As shown in the inset of Fig. 6, at high temperature heat capacity approaches the Dulong-Petit value of 3NR (124.71 J mol$^{-1} $ K$^{-1}$).

From obtained $A$ and $\gamma$, the Kadowaki-Woods (KW) ratio $A/\gamma^{2}$ = 3.43(6)$\times$10$^{-6}$ $\mu\Omega$ cm mol$^{2}$ K$^{2}$ mJ$^{-2}$. This is somewhat smaller than value for KNi$_{2}$Se$_{2}$ polycrystals (1.2 $\times$10$^{-5}$ $\mu\Omega $ cm mol$^{2}$ K$^{2}$ mJ$^{-2}$),\cite{Neilson2} and from the universal value observed in strongly correlated heavy fermion systems (1$\times$10$^{-5}$ $\mu\Omega$ cm mol$^{2}$ K$^{2}$ mJ$^{-2}$).\cite{Tsujii} However, it is still larger than that in many intermediate valence Yb-based and several Ce-based compounds with large $\gamma$ ($A/\gamma^{2}\sim$ 0.4$\times$10$^{-6}$ $\mu\Omega$ cm mol$^{2}$ K$^{2}$ mJ$^{-2}$).\cite{Tsujii} On the other hand, from fitted temperature-independent susceptibility $\chi_{0}$ (1.87(5)$\times$10$^{-3}$ and 4.69(4)$\times$10$^{-3}$ emu mol f.u.$^{-1}$ Oe$^{-1}$ for H$\Vert$ab and H$\Vert$c) and the relation of $\chi_{Landau}\approx-1/3\chi_{Pauli}$ when ignoring the core and orbital diamagnetism, $\chi_{Pauli}\approx$ 2.81$\times$10$^{-3}$ and 7.04$\times$10$^{-3}$ emu mol f.u.$^{-1}$ Oe$^{-1}$ for H$\Vert$ab and H$\Vert$c. The estimate of the Wilson's ratio ($R_{W}=\frac{\pi^2k_{B}^2}{3\mu_{B}^2}\frac{\chi_{Pauli}}{\gamma}$) gives large values $R_{W}$ = 4.26 and 10.68 for H$\Vert$ab and H$\Vert$c, which is much larger than that in a noninteracting Fermi liquid, where $R_{W}$ is expected to be close to 1. This value is also larger than derived from polycrystalline ($R_{W}$ = 1.7).\cite{Neilson2} Such large values have been found in the heavy Fermi liquids ($R_{W}$ =1 - 6).\cite{Delong} Large $R_{W}$ values also occur in the system with a magnetic instability or strong exchange enhanced paramagnetic state.\cite{Julian} All this implies heavy fermion-like ground state in KNi$_{2}$Se$_{2}$ crystals at low temperature.

\section{Conclusion}
In summary, we synthesized K$_{0.95}$Ni$_{1.86}$Se$_{2}$ single crystals using self-flux method. Different from polycrystals, there are small K and Ni deficiencies, similar to iron based counterpart K$_{x}$Fe$_{2-y}$Se$_{2}$. Resistivity measurement indicates that the K$_{0.95}$Ni$_{1.86}$Se$_{2}$ single crystal does not exhibit a superconducting transition down to 0.3 K, in contrast to polycrystals. Therefore, superconducting state is rather sensitive to K and Ni stoichiometry similar to KFe$_{2}$Se$_{2}$. Our results suggest a Fermi surface reconstruction below 30 K, corresponding to the transition from local CDW state to heavy fermion state at low temperature.

\section{Acknowledgement}
Work at Brookhaven is supported by the U.S. DOE under Contract No. DE-AC02-98CH10886 and in part by the Center for Emergent Superconductivity, an Energy Frontier Research Center funded by the U.S. DOE, Office for Basic Energy Science (H. L. and C. P). Work at the National High Magnetic Field Laboratory is supported by the DOE NNSA DEFG52-10NA29659 (D.G.), by the NSF Cooperative Agreement No. DMR-0654118, and by the state of Florida.

\end{document}